\documentclass[a4paper,onecolumn]{agn6}
\usepackage{natbib}
\usepackage{graphicx}
\usepackage[a4paper]{hyperref}
\idline{}{1}

\def\gsimeq
{\hbox{\raise0.5ex\hbox{$>\lower1.06ex\hbox{$\kern-1.07em{\sim}$}$}}}
\def\lsimeq
{\hbox{\raise0.5ex\hbox{$<\lower1.06ex\hbox{$\kern-1.07em{\sim}$}$}}}

\def\hard{$_{2-10 keV}$}
\def\aap{{A\&A}}

\def\aj{{AJ}}

\def\apj{{ApJ}}
\def\apjl{{ApJ}}
\def\apjs{{ApJS}}

\def\deg{$^{\circ}$}

\def\percm2{cm$^{-2}$}

\def\cha{{\it Chandra}~}
\def\xmm{{\it XMM-Newton}~}

\def\ltsima{$\; \buildrel < \over \sim \;$}
\def\simlt{\lower.5ex\hbox{\ltsima}}
\def\gtsima{$\; \buildrel > \over \sim \;$}
\def\simgt{\lower.5ex\hbox{\gtsima}}

\begin{document}
   \title{Multi-wavelength study of a sample of nearby Seyfert galaxies
%\thanks{this is a place for a title footnote}
}

   \author{F. Panessa \inst{1,2}, 
   L. Bassani, M. Cappi, M. Dadina \inst{1},
        and   K. Iwasawa \inst{3}\fnmsep
%\thanks{this is a place for placing a footnote in the author field }
}

   \institute{IASF-CNR, via Gobetti  101, 40129 Bologna,
    Italy \email{panessa@bo.iasf.cnr.it} 
              \and  Center for Astrophysics, 60 Garden Street,
	      Cambridge MA
              \and Institute of Astronomy, Madingley Road, 
		Cambridge, UK}

   \abstract{
One of the least biased complete Seyfert sample available 
to date has been considered to perform an X-ray 
survey using the capabilities of XMM-Newton and Chandra satellites.
The X-ray nuclear luminosities obtained have been used to 
perform a multi-wavelength analysis as a powerfull tool in the identification of
heavily obscured objects.
The new X-ray data, together with highest quality nuclear fluxes
available from the radio to the optical band, allow us to obtain
nuclear Spectral Energy Distributions (SEDs) for type 1 and 
type 2 objects in our sample. The results obtained from the analysis of the 
SED have been discussed in the framework of Unified Models. 
   }
   \authorrunning{F. Panessa et al.}
   \titlerunning{Multi-wavelength nearby Seyfert galaxies}
   \maketitle
%
%________________________________________________________________

\section{Introduction}

Active Galactic Nuclei (AGNs) produce a large amount of energy in small spatial scales and
radiate their energy over a wide range of energies, from gamma-rays 
and X-rays through the ultraviolet, optical and infrared spectral 
regions to the far-infrared and radio-frequency regions. 
X-ray emission is arguably the most important aspect of understanding AGNs, 
because of its unambiguous association
with genuine nuclear activity, and its important diagnostic capabilities
for studying accretion mechanisms.

Nonetheless, the study of the continuum emission of Seyfert galaxies over a
broad range of frequencies is an important tool to 
understand/discriminate the energy output in 
different wavebands, as well as a mean to
distinguish galaxies of different activity classes
and to test Unified Models which ascribe the differences between type 1 and type 2 Seyferts as 
merely due to orientation effects rather than fundamental physical differences.
Previous works, however, were limited since
they concentrated nearly exclusively on high-luminosity AGNs
(L$_{bol}$ $\geq$ 10$^{44}$ erg/s) and very little data exist on 
the spectral properties of low-luminosity
sources (low-luminosity Seyfert galaxies and Liners, with 
L$_{bol}$ $\sim$ 10$^{41}$-10$^{43}$ erg/s). 
To date, only a handful of objects have been adequately studied
with multi-wavelength observations obtained with sufficient angular
resolution in order to detect the nuclear emission
and separate it from the host galaxy background.

We have performed an extensive X-ray study of a complete sample of AGNs unbiased against 
low-luminosity sources using the capabilities of the \cha and \xmm
observatories. We obtained high quality nuclear SEDs for type 1 and type 2 objects
in our sample from the radio to the hard X-ray band, 
using the combination of high angular resolution capabilities and 
unprecedented sensitivity of the new generation telescopes.
We refer to forthcoming publications for details on the 
results here presented (Panessa et al. in preparation, Cappi et al. in
preparation).

\section{The sample}

The sample of Seyfert galaxies has been derived from the Palomar 
optical spectroscopic survey of nearby galaxies (Ho, Filippenko, \& Sargent 
1995). From this survey, high-quality spectra of 486 bright 
($B_T\,\leq$ 12.5 mag), northern ($\delta\,>$ 0\deg) galaxies 
have been taken and a comprehensive, homogeneous catalog of spectral 
classifications of all galaxies have been obtained (Ho et al. 1997). 
The Palomar survey is complete to $B_T$ = 12.0 mag and 80\% complete 
to $B_T$ = 12.5 mag (Sandage, Tammann, \& Yahil 1979).  
From the entire  Ho et al. (1997) sample we have extracted all
Seyfert galaxies\footnote{In a few cases the classification assignment is ambiguous, 
and more than one is given. Objects in which the Seyfert classification
was not dominant but still present (L2/S2, H/S2 or T2/S2) have been 
included in the present sample. Hereafter we refer to these objects
as to "Seyfert dim objects".}.
The total sample of 60 Seyfert galaxies includes 39 type~2 
(type 2, 1.8 and 1.9), 13 type~1 (type 1.0, 1.2, 1.5) and 8 
"Seyfert dim objects".

\section{Chandra and XMM-Newton data analysis}

An homogeneous and standard X-ray data analysis has been carried out on
our selected Seyfert sample using \cha and \xmm observations for 39
objects of the sample with 22 objects having 
observations with both satellites\footnote{To complement the X-ray informations on 
the whole sample, a search in the literature for observations with previous satellites 
(operating in the 2-10 keV energy range) has also been carried out. 
{\it ASCA} observations have been found 
for 8 objects. At the end, 47 sources out of 60 have X-ray data available.}.

\begin{figure}	
\begin{center}
\parbox{16cm}{
\includegraphics[width=0.55\textwidth,height=0.34\textheight,angle=0]{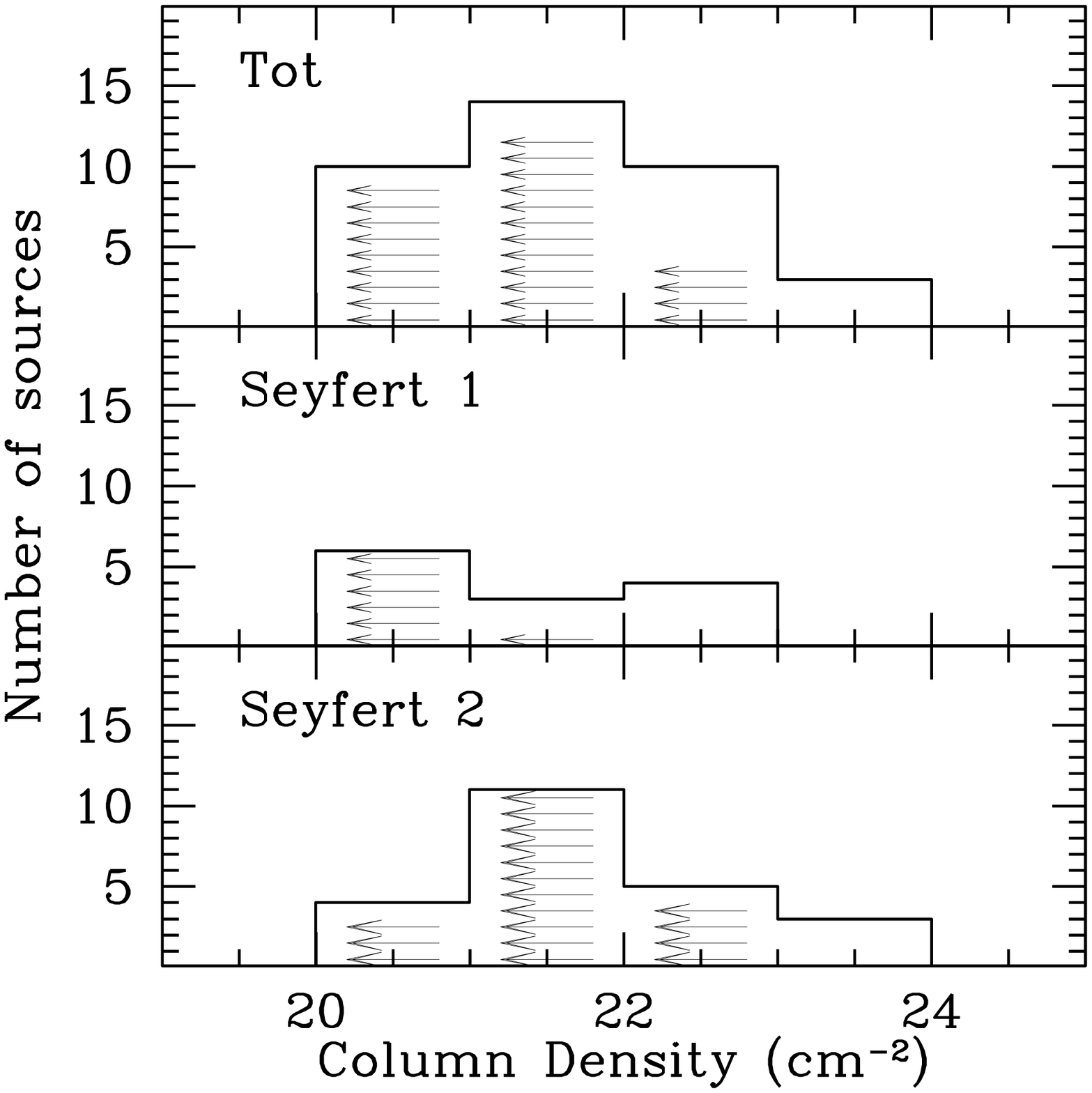}
\includegraphics[width=0.55\textwidth,height=0.34\textheight,angle=0]{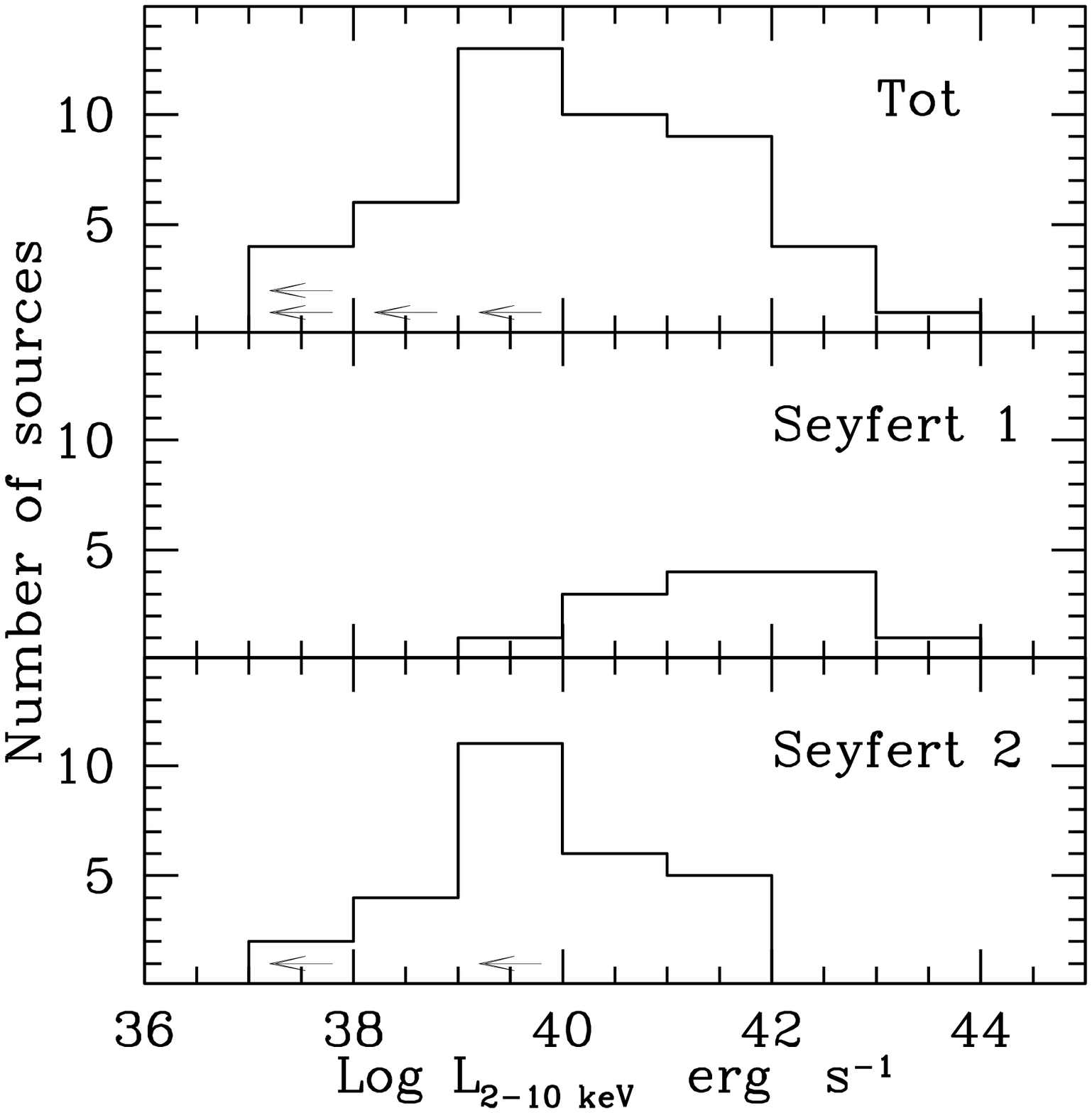}}
\caption{Observed column density distribution (left panel). Observed 2-10 keV luminosity distribution (right panel).
Upper and lower limits have been indicated with arrows.}
\label{nh}
\end{center}
\end{figure}

An atlas of \cha and \xmm images and spectra has been produced
in the 0.3-10 keV energy band. 
The results obtained indicate a high detection rate ($\sim$ 95\%)
of active nuclei, characterized, in $\sim$ 60\% of the objects, also by
the presence of nearby off-nuclear sources and/or 
in $\sim$ 35\% of the objects, of diffuse emission.
Altogether these results demonstrate that
high spatial resolution is fundamental for this type of studies in order
to isolate nuclear emission from other X-ray emitting 
components of the host galaxy.
%A high detection rate of X-ray nuclei in our sample has been found, 
%particularly for type 1 objects which all show a point-like objects, 
%in contrast with type 2 Seyferts which show different morphologies.
Spectral analysis has been performed in order to first identify the
underlying continuum when possible, then additional components and
features have been included to best reproduce the data. 
The distribution of spectral parameters, in particular for type 1 objects,
are found to be within the range of values observed in bright AGNs.
In figure \ref{nh}, left panel, the observed distribution
of column densities for the total sample is shown ranging from the typical Galactic
values to very high absorptions (i.e. $\sim$ 10$^{23}$ cm$^{-2}$).
Nearly 30\% of type 1 Seyfert galaxies are characterized by a significant 
amount of absorption (i.e. $\geq$ 10$^{22}$ cm$^{-2}$) that could be
ascribed to ionized material and/or dense gas clouds crossing the line of sight. 
The distribution of the observed column densities for type 2 Seyfert galaxies here
apparently deviates from past results showing mostly mildly absorbed 
objects. However, it is well known that Compton thick
sources (with N$_{H}$ $>$ 10$^{24}$ cm$^{-2}$) may appear as objects
with little absorption (Risaliti et al. 1999).
%it must be taken into account that
%X-rays above a few keV can penetrate the absorbing material while
%for values $>$ 10$^{24}$ cm$^{-2}$ the photoelectric cutoff (if any) in the observed 
%spectrum does not provide information on the real column density 
%absorbing the primary X-ray source, and so the galaxy may be erroneously 
%classified as a low-absorption object.
In this case we need to use independent
tools to assess their true nature (i.e. heavily absorbed vs. intrinsically 
unabsorbed).

The right panel of figure \ref{nh} shows the observed 2-10 keV luminosity 
distribution for the total sample: a wide range
of luminosities is covered, from objects
with luminosities comparable to those of binary systems 
(i.e. L\hard $\sim$ 10$^{37}$ erg/s) to those with luminosities 
typical of bright AGNs (i.e. L\hard $\sim$ 10$^{43}$ erg/s).
It has been shown in previous surveys that Seyfert 2 galaxies are
generally weaker than their type 1 counterparts. The distribution
of type 1 (middle panel) and type 2 (lower panel) objects
of our sample confirms, at a first glance, this evidence. A Kolmogorov-Smirnov
test to compare the two distributions confirms that they are different at 
$>$99.9\% confidence level.
We will address below whether this difference could be ascribed 
to absorption effects or whether it is an intrinsic property of these sources.

\section{Diagnostics for heavy absorption}

The issues related to the distribution of the
absorption among objects in our sample have been
investigated taking advantage of some powerful diagnostic tools
in order to unveil the presence of obscuration in Seyfert galaxies.

\begin{figure}	
\begin{center}
\parbox{16cm}{
\includegraphics[width=0.55\textwidth,height=0.34\textheight,angle=0]{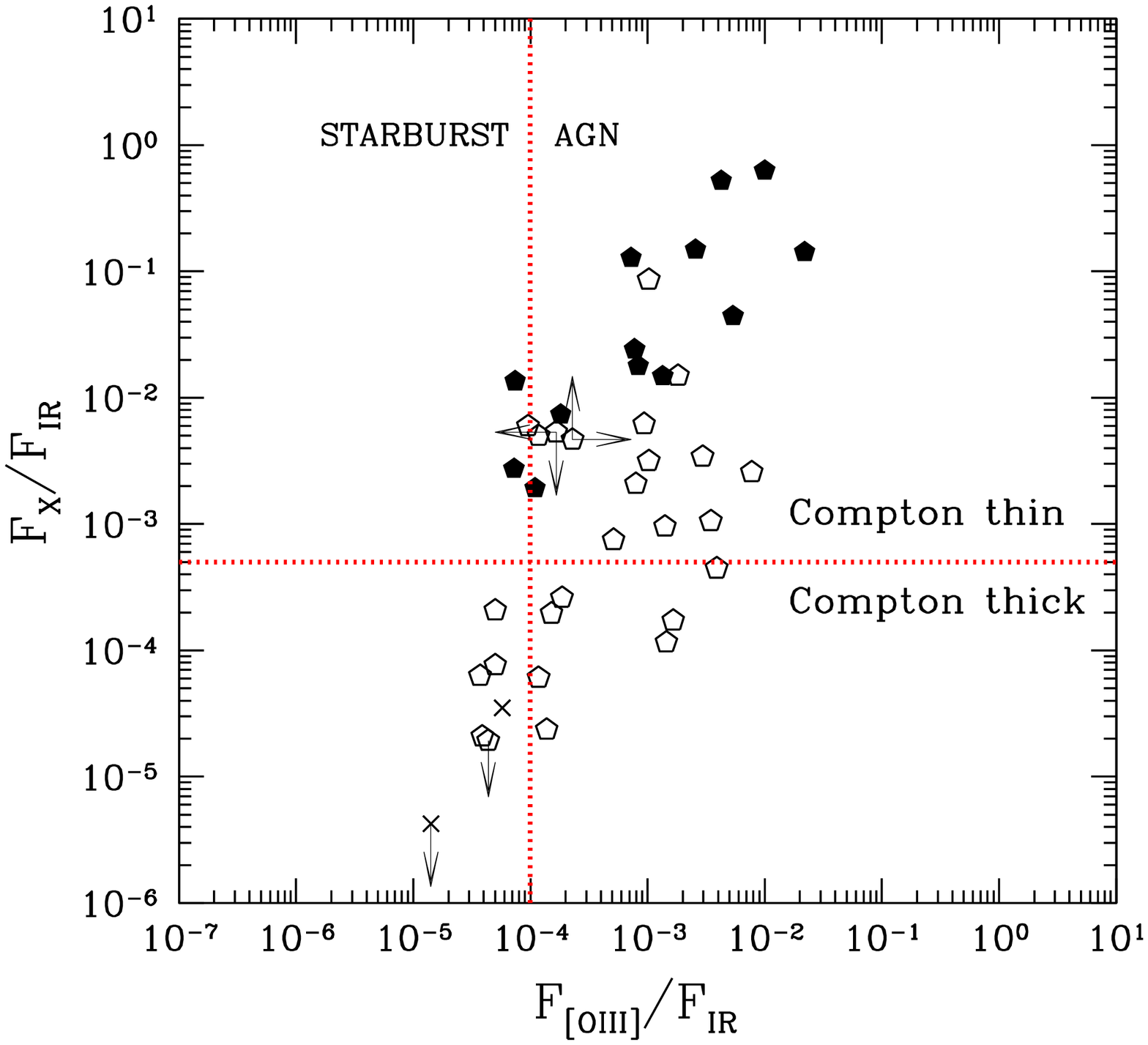}
\includegraphics[width=0.55\textwidth,height=0.34\textheight,angle=0]{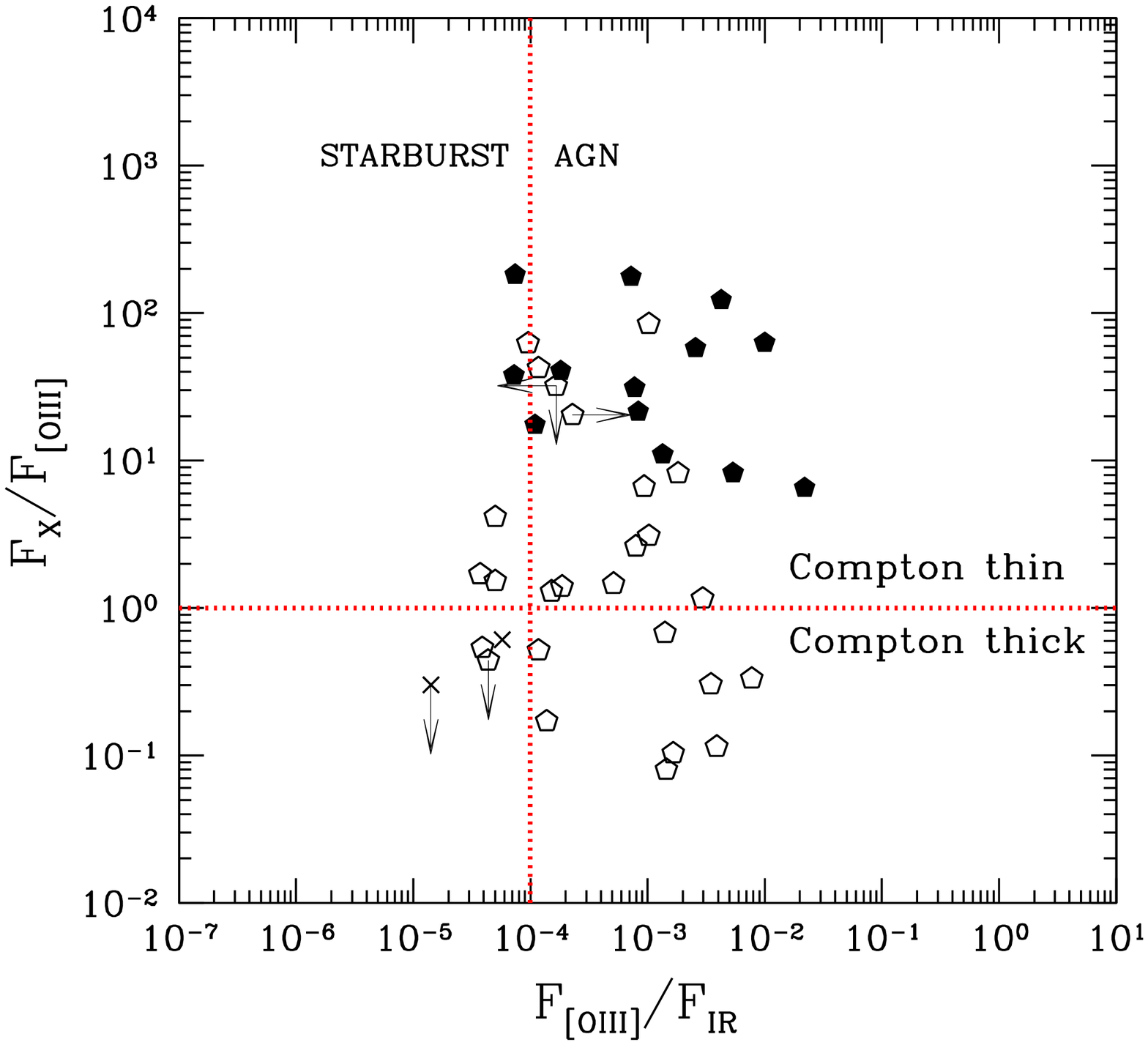}}
\caption{Top panel: F$_{X}$/F$_{IR}$ vs. F$_{[OIII]}$/F$_{IR}$ for the sources of the
sample. Bottom panel: F$_{X}$/F$_{[OIII]}$ vs. F$_{[OIII]}$/F$_{IR}$ for the sources of the
sample. Compton thin, Compton thick and starburst regions have been separated by dashed lines.
Type 1 objects are plotted as filled polygons, type 2 as empty polygons and transition objects
as crosses.}
\label{dia}
\end{center}
\end{figure}	

\subsection{Flux diagnostic diagrams}

Measuring the X-ray luminosity and compare it with an 
isotropic indicator of the 
intrinsic brightness of the source offers an indirect method 
for evaluating the true amount of absorption.
In particular, the [OIII]$\lambda$5007 
flux is considered a good isotropic indicator because it is produced in the  
Narrow Line Region (Maiolino \& Rieke 1995; Risaliti et al. 1999; 
Bassani et al. 1999). Also the Far-Infrared emission seems to be produced 
over a larger region than that of the molecular torus and, thus, has been 
used as an isotropic indicator too 
(Mulchaey et al. 1994, Mas-Hesse et al. 1994).

The F$_{2-10 keV}$/F$_{IR}$, F$_{2-10 keV}$/F$_{[OIII]}$ and F$_{[OIII]}$/F$_{IR}$
ratios, as plotted in the diagrams of figure \ref{dia}, can provide an independent way to establish
which is the dominant component between AGN or starburst and 
at the same time they are a powerful tool in the detection of 
Compton thick sources when an X-ray spectral analysis is not sufficient 
(Panessa \& Bassani 2002).  
The use of such diagrams allows us to recognize a few of Compton thick 
candidates that the X-ray spectral analysis has not been able to 
retrieve. Moreover, a small fraction of objects occupies the Starburst region
pointing to stellar processes as the underlying agent responsible
for the activity.
 
\subsection{The N$_{H}$ vs F\hard/F$_{[OIII]}$ diagram}          

\begin{figure}	
\centerline{\includegraphics[width=8cm]{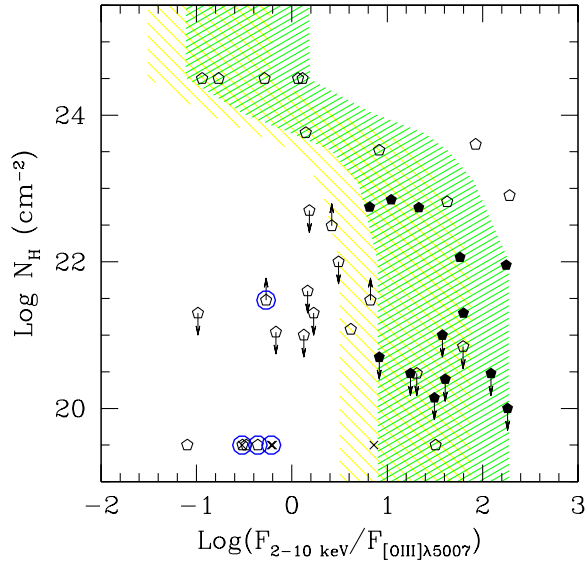}}
\caption{Diagram of the absorbing column density N$_{H}$ versus
the ratio between the observed 2-10 keV flux and the reddening corrected 
[OIII] flux. Filled polygons are type 1 Seyfert, open polygons are type 2 Seyfert,
and 'Seyfert dim' objects are indicated as crosses. The shaded region (lower left to
upper right) indicates the expected correlation by assuming that
L$_{2-10keV}$ is absorbed by the N$_{H}$ reported on the Y-axis, starting from
the average F\hard/F$_{[OIII]}$ ratio observed in type 1 Seyfert galaxies
and by assuming a 1\% reflected component. Also the shaded region 
(upper left to lower right) obtained by Maiolino et al. (1998) is shown.
Possible starburst candidates have been marked with circles.
}
\label{nhcorr}
\end{figure}

The effect of the column density is to decrease the 
F\hard/F$_{[OIII]}$ ratio with respect to Seyfert 1 galaxies.
The reduction is at most by a factor $\sim$ 5 when N$_{H}$ is less than
a few times 10$^{23}$ cm$^{-2}$, and by about two orders of 
magnitude when $>$ 10$^{24}$ cm$^{-2}$. 
In figure \ref{nhcorr} the column densities measured from our 
X-ray spectra have been plotted versus the logarithm of the F\hard/F$_{[OIII]}$ ratio.
The shaded region (lower left to
upper right) indicates the expected correlation by assuming that
F$_{2-10keV}$ is absorbed by the N$_{H}$ reported on the Y-axis, starting from
the average F\hard/F$_{[OIII]}$ ratio observed in type 1 Seyfert galaxies
and assuming a 1\% reflected component.
The width of the shaded region has been drawn considering the lower and higher
F\hard/F$_{[OIII]}$ ratios of the type 1 Seyferts of our sample.
The shaded region (upper left to lower right) obtained by Maiolino et al. (1998) has also been reported
for comparison: it is evident that our and their
results are substantially similar confirming
the relation between the X-ray versus [OIII] flux ratio and
column density\footnote{ 
There are 8 objects in our sample for which an estimate 
of the column density is not available, 
they have been plotted assuming that they are absorbed only by a 
Galactic column density of 5 $\times$ 10$^{19}$ cm$^{-2}$.}.

Compton thick objects should occupy the high 
N$_{H}$ - low F\hard/F$_{[OIII]}$ region of the predicted 
distribution. 
A large fraction of objects follow the expected distribution,
while $\sim$ 30\% of the sources are clearly out of the correlation
in the low N$_{H}$ - low F\hard/F$_{[OIII]}$ part of the diagram.
These objects would lie in the correlation if they were
absorbed by a column density higher than few times 10$^{23}$ cm$^{-2}$.
Among these objects we find five of them
which have been found in the Starburst region from the flux diagnostic diagrams 
(figure \ref{dia}): all of them have been marked with a 
ring in figure \ref{nhcorr} to highlight their difference
from the other Seyfert galaxies. Their different behaviour
is also confirmed by our X-ray analysis which indicates that
the nuclear cores of these objects appear to be very faint or absent. 
Since in these objects it is not even clear that there is
an AGN, we prefer to be conservative and treat them as a separate
class, however these sources 
may also be affected by large absorption and turned out to be 
Compton thick if deeper X-ray measurements were to be done.

Sources having a F\hard/F$_{[OIII]}$ ratio smaller than $<$ 1 are 
our most likely Compton thick candidates. 
 
It is worth noting a few anomalous cases in
figure \ref{nhcorr}: (i) type 1 objects (filled polygons) having the
absorbing column greater than 10$^{22}$ cm$^{-2}$ 
and (ii) type 2 objects (open polygons) which lie in the region of 
high F\hard/F$_{[OIII]}$ - low N$_{H}$ occupied by type 1 objects.

\subsection{Corrected column density and X-ray luminosity distribution}

After considering multi-wavelength diagnostics it turns out that the fraction of
absorbed objects is largely underestimated if the column density measurements are
based only on the 2-10 keV spectrum. As a consequence, the observed column density
and 2-10 keV luminosity distributions should be corrected taking into account our
Compton thick candidates. Those tools applied to our sample sources, have revealed
that the fraction of objects which may be affected by Compton thick   obscuration
ranges from 20\% up to 50\% in agreement  with previous  estimates available for a
flux-limited sample (Risaliti et al. 1999). With the present work we are able to
probe much lower luminosities and still find that the fraction of absorbed objects
remains significantly high.

As a consequence, the dichotomy often observed in the luminosity of type 1 and type 2
AGNs is, therefore, mainly to be ascribed to the presence of heavy absorption in type 2
objects. In fact, correcting for the absorption, 
we find a flat distribution of N$_{H}$ and no significant difference in 
luminosity of type 1 and type 2 Seyfert galaxies. Overall, the results obtained here are  therefore in agreement with the
predictions of unified models, except for a few ($\sim$ 10\%)  particular cases which do
not fit easily into the standard picture:  (i) bona-fide type 2 objects which have no 
evidence for absorption in excess to the Galactic value and (ii) type 1 Seyferts which
have column densities of $\sim$ 10$^{22}$-10$^{23}$ cm$^{-2}$, i.e. more typical of type
2 objects. Another small fraction ($\sim$ 10\%) of objects in our sample, characterized
by luminosities lower than $\sim$ 10$^{38}$ erg s$^{-1}$, is of ambiguous nature, i.e.
it is not clear whether a nuclear starburst  or an active nucleus or a combination of
both are responsible for the observed emission.

\section{Nuclear Spectral Energy Distributions}

Recently, multi-wavelength surveys from the radio to the
Infrared band have been performed on sub-samples
of the Ho et al. (1997) sample of Seyfert galaxies
(Ho \& Ulvestad 2001, Alonso-Herrero et al. 2003, Ho \& Peng 2001). Taking
advantage of these extensive data sets, it is thus possible
to consider here the highest quality nuclear fluxes and therefore 
assemble the SEDs from radio ($\nu\approx10^8$ Hz) to hard X-rays 
($\nu\approx10^{19}$ Hz) for a significant fraction of the objects
in our sample. 
%obtained from literature and X-ray data as described in chapter 4.
High resolution 2-10 keV observed fluxes as obtained from the X-ray analysis
have been used in assembling the SEDs.

\begin{figure}	
\begin{center}
\parbox{16cm}{
\includegraphics[width=0.55\textwidth,height=0.34\textheight,angle=0]{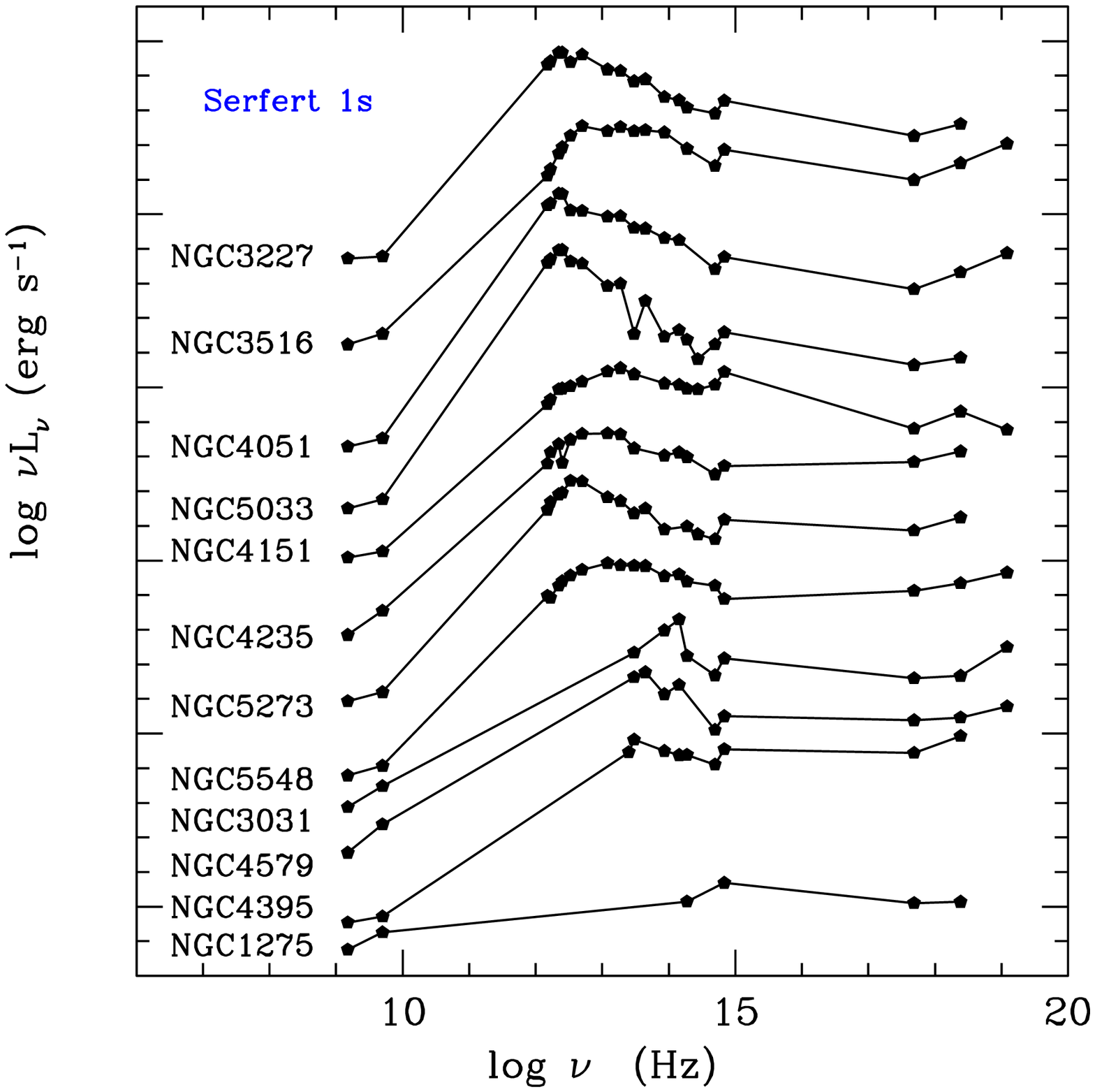}
\includegraphics[width=0.55\textwidth,height=0.34\textheight,angle=0]{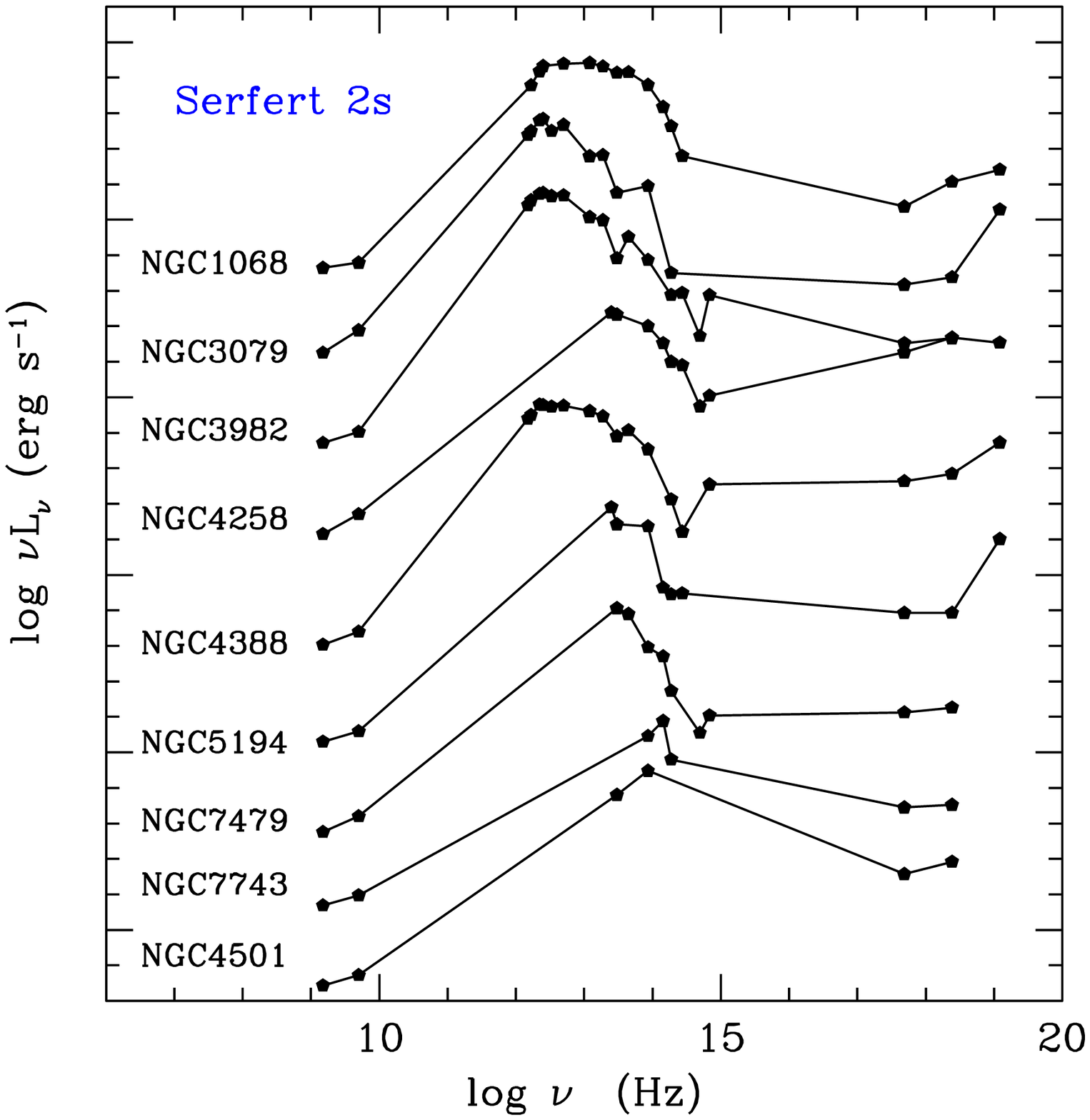}}
\caption{Individual SEDs, separated vertically by arbitrary
constants for clarity. The galaxy name is shown on the left of each SED. Top
panel: Seyfert~1's. Bottom panel: Seyfert~2's.}
\label{sedtot}
\end{center}
\end{figure}

The best quality individual SEDs of type 1 and type 2 Seyfert galaxies
are presented in figure \ref{sedtot},
shifted vertically by arbitrary constants, for clarity. The data from the radio 
to the hard X-rays have been drawn as
filled dots. A straight line has also been drawn to connect
the data points and better compare the SEDs\footnote{Note that in 
the millimeter and in the UV region
this line does not represent the real SED.}.

The SEDs reported in figure~\ref{sedtot} indicate the same
behaviour in both type 1 and type 2 objects going 
from the radio to the far infrared band: clearly in each 
object the flux increases going from 6 cm to 200 $\mu$m.
Figure~\ref{sedtot} also shows that
the infrared energy band is very well described over $\sim$ 3 decades of frequencies
from the far-infrared to the near infrared band.
It is interesting to note that 
%when two different data sets are combined for both type~1 
%and type~2 objects of our sample, as in figure \ref{sedtot}, 
type~1 sources show a flatter spectrum all over the infrared waveband
with respect to type 2 objects in which $\nu$L$_{\nu}$ 
decreases with increasing frequency. Moreover, in type~1 objects 
the spectra appear to be flat enough as to meet the observed 
emission at X-ray frequencies, while in type~2 objects
it is likely that absorption has the effect of reducing 
the emission also in the optical and X-ray wavebands.

\begin{figure}	
\centerline{\includegraphics[width=8cm]{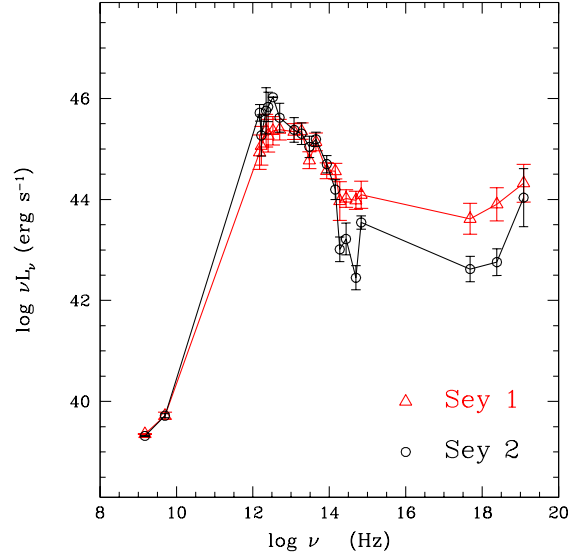}}
\caption{Comparison between the average SED of Seyfert~1 and Seyfert~2. The two
SEDs have been normalized at 6 cm. The error bars are the standard deviation of the average.}
\label{ave}
\end{figure}

The average continuum spectral distribution from radio to hard X-rays for type 1 and
type 2 Seyfert have been plotted  in figure \ref{ave} normalized to the flux at 6 cm.
In both types the far and mid infrared  components appear to dominate the energy output,
while the  two mean SEDs differ from the near-infrared band (at 1.6$\mu$m) up to 10 keV,
reconciling at hard X-ray frequencies, namely at 50 keV\footnote{It is worth noting that
the luminosity in the B band ($\nu$ $\sim$ 10$^{15}$ Hz)   has been obtained assuming
$f_{\nu}\,\propto\,\nu^{\alpha_{\rm o}}$, with  $\alpha_{\rm o}\,=\,-1.0$ by Ho \&
Peng~(2001),  which is typical for Seyfert~1 nuclei, however this assumption may not be
valid for type 2 objects.}.

A comparison of the average spectral properties of 
the two classes of Seyferts show that: (i) in both cases the 
far and mid infrared emission appear to dominate the energy output,
(ii) the near infrared, optical and X-ray emission up to 10 keV
are significantly different in the two classes, type 2 objects
showing lower luminosities.
The results obtained are in agreement with the predictions of
unified models in which Seyfert 2s are the
obscured version of Seyfert 1s. 
Detailed models will be applied to our data in order to describe
the spectral shapes and therefore put constraints 
on the main properties of the absorbing medium.

\begin{acknowledgements}
	This work has been carried on during the Ph.D. thesis of F.~Panessa and it has been
	supported by the Consiglio Nazionale delle Ricerce (Italy).
\end{acknowledgements}

\bibliographystyle{aa}

\end{document}